# Nitrogen-Functionalized Graphene Nanoflakes (GNFs:N): Tunable Photoluminescence and Electronic Structures

J. W. Chiou,*,†,Φ Sekhar C. Ray,*,‡,§,Φ S. I. Peng,‡ C. H. Chuang,‡ B. Y. Wang,‡ H. M. Tsai,∥ C. W. Pao,∥ H.-J. Lin,∥ Y. C. Shao,‡ Y. F. Wang,‡ S. C. Chen,‡ W. F. Pong,*,‡ Y. C. Yeh,⊥ C. W. Chen,⊥ L.-C. Chen,# K.-H. Chen,∇ M.-H. Tsai,○ A. Kumar,◆ A. Ganguly,◆ P. Papakonstantinou,◆ H. Yamane,¶ N. Kosugi,¶ T. Regier,□ L. Liu,+ and T. K. Sham+

†Department of Applied Physics, National University of Kaohsiung, Kaohsiung 811, Taiwan
‡Department of Physics, Tamkang University, Tamsui 251, Taiwan
§School of Physics, DST/NRF Centre of Excellence in Strong Materials and Materials Physics Research Institute (MPRI); University of the Witwatersrand, P/Bag 3, WITS 2050, Johannesburg, South Africa
∥National Synchrotron Radiation Research Center, Hsinchu 300, Taiwan
⊥Department of Materials Science and Engineering and #Center for Condensed Matter Sciences, National Taiwan University, Taipei 106, Taiwan
∇Institute of Atomic and Molecular Sciences, Academia Sinica, Taipei 106, Taiwan
○Department of Physics, National Sun Yat-Sen University, Kaohsiung 804, Taiwan
◆Engineering Research Institute (ERI), School of Electrical and Mechanical Engineering, University of Ulster at Jordanstown, Newtownabbey, Co. Antrim BT37 OQB, N. Ireland
¶Institute for Molecular Science, Okazaki, Japan
□Canadian Light Source, Inc., Saskatchewan S7N 0X4, Canada
+Department of Chemistry, University of Western Ontario, London N6A 5B7, Canada

**Ⓢ** Supporting Information

**ABSTRACT:** This study investigates the strong photoluminescence (PL) and X-ray excited optical luminescence observed in nitrogen-functionalized 2D graphene nanoflakes (GNFs:N), which arise from the significantly enhanced density of states in the region of $\pi$ states and the gap between $\pi$ and $\pi^*$ states. The increase in the number of the $sp^2$ clusters in the form of pyridine-like N−C, graphite-N-like, and the C=O bonding and the resonant energy transfer from the N and O atoms to the $sp^2$ clusters were found to be responsible for the blue shift and the enhancement of the main PL emission feature. The enhanced PL is strongly related to the induced changes of the electronic structures and bonding properties, which were revealed by the X-ray absorption near-edge structure, X-ray emission spectroscopy, and resonance inelastic X-ray scattering. The study demonstrates that PL emission can be tailored through appropriate tuning of the nitrogen and oxygen contents in GNFs and pave the way for new optoelectronic devices.

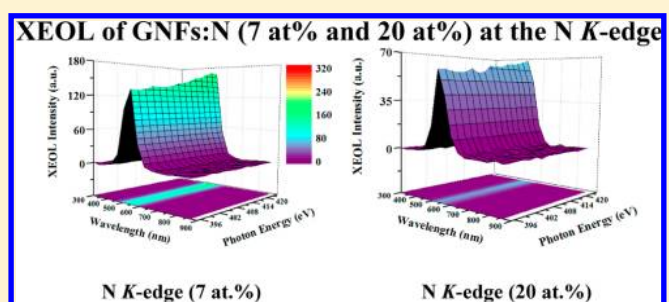

## 1. INTRODUCTION

Graphene is a one-atom-thick material that has been the focus of fundamental and applied solid-state physics research.[1] Novel condensed-matter effects of graphene that arise from its unique 2D energy dispersion and its superior properties make it promising for a wide variety of applications. The lowest unoccupied state and highest occupied state coincide at the $K$ point, and the dispersions of the two corresponding energy bands are essentially linear near the $K$ point, in contrast with the parabolic dispersions near the conduction-band minimum ($E_{CBM}$) and valence-band maximum ($E_{VBM}$) of a semiconductor. The absence of an energy-gap gives rise to a unique set of challenges for its implementation in nanometer scale electronics. For example, it is associated with a substantial leakage current in field-effect transistors,[2] and it renders luminescence, which is thought to be highly unlikely. However, photoluminescence (PL) has been observed in chemically modified graphene whose electronic structure has been modified.[3−6] There are two main methods for modifying the electronic structure of graphene. One is to cut it into ribbons or quantum dots,[7,8] and the other is the chemical or physical treatment with various gases to reduce the connectivity of the $\pi$ electron network.[9,10] Chemically functionalized graphene oxide (GO) or reduced graphene oxide (r-GO) is the most popular chemically derived graphene-based material[10] that exhibits a broad luminescence emission.[4] However, GO or r-GO can be viewed as a graphene modified with oxygen functional groups such as epoxy and hydroxyl, which aggregate on the basal plane, or carbonyl and carboxyl groups at the sheet edges. GO and r-GO have been demonstrated to tune band gap over a large range, suggesting that the functionalization of graphene by oxidation can alter or modify its electronic properties.[11,12] Recent optical studies have suggested that GO exhibits a band gap that can be tuned by controlling the degree of oxidation or reduction; the intensities of the two PL features were found to vary upon reduction. This effect was argued to be correlated with the electron–hole recombination in very small $sp^2$ clusters within the carbon–oxygen $sp^3$ matrix and the radiative recombination from disorder-induced localized states in the $\pi$–$\pi^*$ gap.[5,13,14] In contrast, a recent study of the chemically modified Mn-bonded r-GO suggested that the optical properties of the Mn ions could be altered and resonance energy can be transferred from the Mn ions to the $sp^2$ carbon clusters, yielding intense long-wavelength PL emission.[15] The hydrogenation[16] and N-doping[17] of graphene can also modulate the $sp^2$- to $sp^3$-bonding ratio and the electronic properties, which can be used to tune its band gap and optical properties. The PL mechanism in GO/r-GO or functionalized graphene is controversial. Therefore, a direct atom-specific spectroscopic observation to elucidate the effects of the N-doping or GO matrix on the PL mechanism, electronic structures, and bonding properties will be extremely desirable, and the understanding of the origin of light emission and the enhancement of emission in N-doped graphene would be very valuable. This work studies the relationships between PL and the electronic structures and bonding properties of graphene nanoflakes (GNFs) with nitrogen functionalization (GNFs:N) using nitrogen plasma. The combination of C K-edge X-ray absorption near-edge structure (XANES) and C $K_\alpha$ X-ray emission spectroscopy (XES) of GNFs:N and GNFs reveals that they have a metallic character similar to that of highly oriented pyrolytic graphite (HOPG). Resonant inelastic X-ray scattering (RIXS) indicates an increase in the number of nitrogen- and oxygen-induced localized states in the $\pi$ region, or in between $\pi$ and $\pi^*$ regions in GNFs:N. PL emission changes drastically upon the incorporation of nitrogen, which alters the electronic structures in the GNFs matrix. The character of PL emission could be tuned by changing the $sp^2$ domain within the GNF structure by the formation of pyridinic- and graphitic-N structures and the C=O bonding with various at % of N. The results of X-ray-excited optical luminescence (XEOL) further demonstrate that the incident photon energies at the N and O K-edge are responsible for most of the XEOL intensity at ~520 nm, unlike that at the C K-edge, clearly indicating that N and O atoms play an important role to transfer resonance energy to the $sp^2$ clusters, which yields intense short-wavelength emission in GNFs:N, as observed in the PL spectra.

## 2. EXPERIMENTAL SECTION

GNFs and GNFs:N with a thickness <1250 nm were deposited on the Si(100) substrate without a catalyst by microwave plasma-enhanced chemical vapor deposition and subsequently functionalized with RF-plasma glow discharge using nitrogen plasma at different sccm by the electron cyclotron resonance process at room temperature for 5 min. To explore the origin of light emission of this system, we have performed the PL and XEOL spectroscopy. The PL spectra were obtained using a Hitachi F4500 fluorescence spectrophotometer at an excitation wavelength of 400 nm, whereas the N, C, and O K-edge XEOL were measured at the undulator-based, spherical grating monochromator 11ID-1-beamline at the Canadian Light Source, Saskatchewan. The XEOL spectra were collected using a dispersive spectrometer (QE65000, Ocean Optics). PL yields were measured by collecting total (zero order, 200−950 nm) and wavelength-selected luminescence as the excitation photon energy was tuned across the absorption. The electronic structures and bonding properties were studied using XANES, XES, RIXS, and X-ray photoelectron spectroscopy (XPS). The XANES spectra were obtained using the high-energy spherical grating monochromator 20A-beamline at the National Synchrotron Radiation Research Center (NSRRC), Hsinchu, Taiwan and the soft X-ray in-vacuum undulator BL3U-beamline with the valid-line-spacing plane grating monochromator at the Ultraviolent Synchrotron Orbital Radiation (UVSOR) facility, Institute for Molecular Science (IMS), Okazaki, Japan. C $K_\alpha$ XES and RIXS spectra were obtained using the transmission-grating XES/RIXS spectrometer at the BL3U-beamline at the UVSOR facility, IMS. The XPS spectra were obtained using the low-energy spherical grating monochromator 08A-beamline at NSRRC. The microstructure and surface morphology were studied using Raman spectroscopy and scanning electron microscopy (SEM). Raman spectra were performed using excitation energy of 633 nm (1.96 eV) and a spectral resolution of better than 2 cm$^{-1}$, whereas SEM was examined using a JEOL JSM-6700F scanning electron microscope. In addition, the electron field emissions (EFEs) were measured using a Keithley power supply.

## 3. RESULTS AND DISCUSSION

Figure 1a plots the EFE current density ($J$) versus the applied electric field ($E_A$) curves. The insets in Figure 1a present the SEM images of GNFs:N (7 and 20 at %) and pure GNFs. Initially, $J$ decreases upon nitrogenation to 7 at %, as indicated by the curved down-arrow, and then increases at 20 at %, as indicated by the curved upward arrow. The absence of any significant difference between the SEM images reveals a lack of change in surface morphology. The turn-on electric field ($E_{TOE}$) was determined by the Fowler–Nordheim (F–N) plot to be ~9 V/$\mu$m at an emission current density of 0.01 mA/cm$^2$.[18] $E_{TOE}$ increases initially from 9 V/$\mu$m (pure GNFs) to 15 V/$\mu$m for GNF:N (7 at %) and then decreases to 7.5 V/$\mu$m for GNFs:N (20 at %). The EFE of carbon-containing materials depends strongly on the $sp^2$ content and the insulating $sp^3$ matrix in the structure. In the present case, the $sp^2$ content is sufficiently high in GNFs:N with 20 at % N, unlike in pure and 7 at % nitrogenated GNFs, as to be discussed in the following.

Figure 1b presents the Raman spectra of GNFs:N and GNFs in the range 1200−2800 cm$^{-1}$ obtained by laser excitation of wavelength of $\lambda_{ex}$ = 633 nm. The visible Raman spectra of GNFs:N and GNFs include two main peaks G and D in the region 1200−1700 cm$^{-1}$, which depend on the $sp^2$/$sp^3$ ratio of the carbon bonds.[19−21] They also exhibit a 2D band at ~2665 cm$^{-1}$. The G peak is associated with the presence of both olefinic (chains) and aromatic (rings) $sp^2$ structures known as the $E_{2g}$ mode. and the D peak is associated with the aromatic rings only.[19] The 2D band at ~2660 cm$^{-1}$ is related to disorder-induced defects, and its shape is highly sensitive to the number of grapheme layers.[19,20] The spectra of GNFs:N and GNFs show that the D- and G-band maxima are located at ~1340 and ~1582 cm$^{-1}$, respectively, and a split 2D band is located at ~2660 and 2670 cm$^{-1}$, which correspond to two- and three-layer graphenes.[20,21] Additionally, a shoulder at ~1617 cm$^{-1}$, D′, can be attributed to an intravalley process,[22,23] which is characteristic of the structural disorder in a single/few-layer graphene. If enough defects are present in the sample, then a second defect D′ band may appear. No shift of the D band is observed, but a very slight broadening and an increased intensity are observed for the D band, whereas the G band becomes slightly narrower and its intensity is decreased upon nitrogenation, indicating that the graphite-like structures are more disordered than the pyridine-like ones. The D peak/G peak intensity ratio, $I_D/I_G$ (a sp$^2$/sp$^3$ signature), is enhanced substantially upon functionalization with N [GNFs: 0.75 → GNFs:N (7 at %): 1.08 → GNFs:N (20 at %): 1.03], revealing that nitrogenation yields some structural imperfections and the changes of the crystallinity. The incorporation of N into the GNFs structure generates C–N and N–N bonds at the expense of the C–C bonds. However, the C–N vibration modes, which lie between the G and D bands, cannot be identified owing to the insensitivity of Raman excitation to distinguish between the cross sections of C and N atoms.

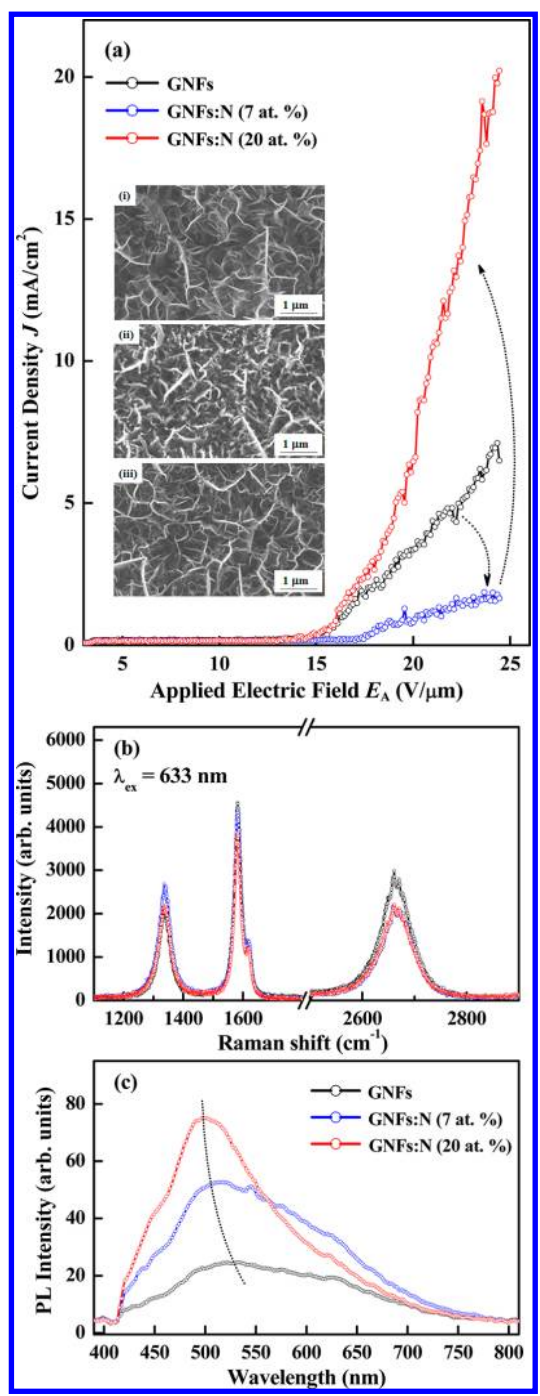

**Figure 1.** (a) Current density ($J$) versus applied electric field ($E_A$) characteristics of the electron field emission of the GNFs:N and GNFs thin films. The inset of panel a presents scanning electron microscopy images: (i) GNFs, (ii) GNFs:N (7 at %), and (iii) GNFs:M (20 at %). (b) Raman spectra. (c) PL spectra of GNFs:N and GNFs.

Figure 1c displays the PL spectra of GNFs:N and GNFs. Because graphene has no band gap, no PL is expected upon the relaxation of charge carriers.[24] Because of the weak inelastic scattering that is associated with the Raman process, PL can be observed only in chemically modified graphene, in which the electronic structures or bonding properties have been modified, as mentioned above.[3−5] Strong PL emission from GNFs:N and weak PL emission from GNFs have been observed under the excitation by 400 nm (~3.1 eV) laser light. Emission across the visible spectral range, 415–800 nm (2.99–1.55 eV), was observed, but the intensity of long-wavelength 650–800 nm PL was always very weak, as shown in Figure 1c. Nitrogen functionalization of GNFs increases the intensity of PL and shifts the main feature toward shorter wavelengths from ~530 nm in GNFs to ~500–510 nm in GNFs:N, as indicated by the dashed line. The intensity of PL is found to increase [GNFs: $7 \times 10^3$ → GNFs:N (7 %): $12 \times 10^3$ → GNFs:N (20 %): $13 \times 10^3$] with the N/O ratio in GNFs:N. Eda et al.[5] and Cuong et al.[25] observed that when as-deposited GO is reduced by exposing it to hydrazine, its PL is enhanced, suggesting that the enhancement of the PL emission was related to N and O or their ratio. The variation of PL shown in Figure 1c suggests a correlation between the contents of N and O or its ratio in the GNFs matrix. The highest PL is observed for the GNFs that contained 20 at % N (and had the highest N/O ratio of ~2.5). The amount of oxygen content in GNFs:N, which was determined by the XPS compositional analysis (Table 1), is presented in Figures S1−S3 of the Supporting Information. PL emission can be tuned by controlling the nature of the sp$^2$ sites in graphitic carbon materials,[26] and so Wen et al. was able to observe the enhancement of PL for graphitic carbon nanotubes (CNTs) upon treatment with nitrogen plasma owing to the increase in the number of sp$^2$ clusters in the CNT structure.[27] These observations are useful in understanding the changes of the electronic structures and bonding properties, and therefore PL, of the nitrogen- or oxygen-treated GNFs. The oxygen content determined by XPS is consistent with the π* and σ* intensities in the O K-edge XANES spectra, which can also be seen in Figure S4 of the Supporting Information.

**Table 1. X-ray Photoelectron Spectroscopy (XPS) compositional analysis**

| thin films | XPS compositional (at %) | | | |
|---|---|---|---|---|
| | C | O | N | N/O ratio |
| GNFs | 96 | 4 | | |
| GNFs:N (7 at %) | 89 | 4 | 7 | 1.7 |
| GNFs:N (20 at %) | 72 | 8 | 20 | 2.5 |

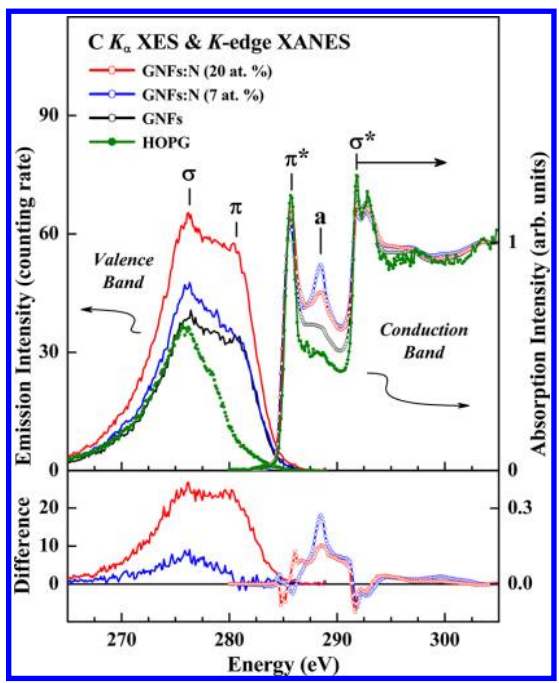

**Figure 2.** Normalized C K-edge XANES and $K_\alpha$ XES spectra of GNFs:N and GNFs. The lower panel shows the GNFs:N spectra subtracted from that of GNFs (right side: XANES, left side: XES).

Figure 2 presents the normalized C K-edge XANES and $K_\alpha$ XES of GNFs:N, GNFs, and reference HOPG. The XANES spectra are normalized to the incident beam intensity, $I_0$, with the area under the spectra in the energy range between 303 and 308 eV fixed (not shown). The XES spectra normalized the intensity to the duration of data acquisition. The XANES and XES spectra probe the unoccupied and occupied C p-like densities of states (DOSs), respectively. Features of $\pi^*$ and $\sigma^*$ character are observed at ~285.5 and 292 eV in the XANES spectra, respectively,[28−30] and the $\pi$ and $\sigma$ bands in the XES spectra of HOPG are observed at 277−284 and 270−277 eV, respectively. The $\pi^*$ feature is typical of the unsaturated C=C or graphitic (sp$^2$) bond above and below the plane, whereas the in-plane $\sigma^*$ feature is typical of the C−C bond. A wide feature centered at ~288 eV, **a**, is observed between the $\pi^*$ and $\sigma^*$ features in the XANES spectra. Ray et al. observed a similar feature in the range (287−291 eV) for the amorphous (a)-C:H(OH) thin films and attributed it to a combination of the C−H $\sigma^*$ band associated with the bonding between the H atom and a diamond-like-bonded C atom with O−C=C, C−OH, and C=O $\pi^*$ bonds.[28] However, this feature was attributed to the interlayer state related to the charge density between the grapheme layers by Pacilé et al.[31] and to the presence of a COOH group or C−H contamination by oxidation by Jeong et al.[32] This feature was also attributed to the C=O $\pi^*$ moiety of the carboxylic group of r-GO by Zhou et al.[19] and by Sun et al.[33] Because this feature is absent in the angle-dependent C K-edge XANES spectra at normal incidence using a scanning transmission X-ray microscopy (STXM), the aforementioned functional groups may lie in the basal plane and are likely to concentrate along the edge of GO.[19] Although the assignment of feature **a** has been controversial, feature **a** of GNFs:N is clearly stronger than those of GNFs and HOPG, suggesting that it is contributed by N- and O-related bonds in GNFs:N. Notably, the intensity of feature **a** dramatically decreases upon vacuum heat treatment at 800 °C, indicating decomposition of these functional groups.[34] Similar bonding structures of N and O with carbon atoms and the sp$^2$ and sp$^3$ hybridized states are observed in C, O, and N 1s core-level XPS and N and O K-edge XANES studies,[21,35,36] which are described in the Supporting Information. Also seen in Figure 2 are the line shapes of the C $K_\alpha$ XES spectra of HOPG, which are known to depend strongly on both incident angle and excitation energy owing to the difference between the intensity distributions of the $\pi$ and $\sigma$ components of the emission.[37−39] Although the XES features in Figure 2 have various intensities, the general spectral line for GNFs:N and GNFs are similar, suggesting that the chemical bonding states are essentially similar. They exhibited a broad main feature centered close to 277.2 eV ($\sigma$ state) and a high-energy shoulder centered near 281.4 eV ($\pi$ state), but the $\pi/\sigma$ ratio of GNFs differs from that of GNFs:N. The height ratios of the $\pi$ and $\sigma$ features are ~0.83 (for pure GNFs), 0.76 (for 7 at % N), and 0.93 (for 20 at % N). This change of the $\pi$ and $\sigma$ ratios can be rationalized as follows: when the amount of nitrogen is changed from 0 to 7 at %, the more electronegative nitrogen tends to withdraw more charge through the $\pi$ bond, and when nitrogen concentration increases from 7 to 20 at %, then it reverses the trend and acts more like a donor than an electron-withdrawing group. This trend is consistent with the field-emission characteristics of the GNFs and GNFs:N films that $E_{TOE}$ increases and $J$ decreases with the increase in $E_A$ when the amount of N increased to 7 at %; then, $E_{TOE}$ decreases and $J$ increases with the increase in $E_A$ when the amount of N increased from 7 at % to 20 at %, as shown in Figure 1a. To examine the effect of nitrogenation on GNFs, the XANES and XES spectra of GNFs are subtracted from those of GNFs:N and are shown in Figure 2 (lower right and left panels, respectively). The integral difference XANES spectra in the energy range of interest is positive, indicating that the number of $\pi^*$ and $\sigma^*$ DOSs increases with nitrogenation in GNFs and further suggesting charge transfer from carbon to the more electronegative nitrogen.[40] The difference XES spectra reveal a quite large positive difference between the intensities obtained from the spectra of GNFs:N (20 at % N) and GNFs, indicating a very large enhancement of the DOSs of $\pi$ and $\sigma$ states in GNFs (20 at % N). Both of the leading edges of the C K-edge XANES and $K_\alpha$ XES spectra have been extrapolated to the baseline in order to determine $E_{CBM}$ and $E_{VBM}$,[41,42] respectively, and the band gap. However, the two extrapolated lines clearly intersect each other, meaning that GNFs:N and GNFs have no band gap similar to the metallic HOPG. According to Chien et al.,[14] the PL emission from GNFs:N and GNFs presented in Figure 1c may arise from interactions between the nanometer-size sp$^2$ clusters (or finite-sized molecular sp$^2$ domains) embedded in the carbon−oxygen (C−OH/C−O−C) sp$^3$ matrix, mainly because of an optical transition that involves the number of substitution induced localized states in the band tail of the $\pi−\pi^*$ gap. Another interpretation is that N and O atoms are incorporated into the GNFs matrix by the formation of functional groups or substitution at C sites, where they serve as energy traps and thereby support the transfer of resonance energy from the N and O atoms to the sp$^2$ clusters. Because N and O atoms can bond to C atoms, forming sp$^2$ clusters, and thus have an active role in GNFs structure, the radiative recombination rate is substantially increased, causing extensive emission from GNFs:N.

To verify directly the increase in the number of substitution-induced localized states in GNFs:N upon nitrogenation, a detailed analysis of the DOSs in the region of π or in the π−π* gap was conducted using RIXS. Figure 3 displays RIXS spectra of GNFs:N (20 at %), GNFs, and HOPG obtained by excitation at various energies. The upper inset of Figure 3 shows the experimental geometry used to obtain the spectra. According to the RIXS theory, the total energy and momentum must be conserved in resonant inelastic scattering during absorption-emission processes.[38,39] Considering highly the texture of GNFs:N, GNFs, and HOPG, Figure 3 reveals that the C $K_\alpha$ RIXS spectra depend strongly on the excitation energy, and this dependence is believed to arise from the resonant inelastic scattering and is related to the band structures of GNFs:N, GNFs, and HOPG. Most of the spectra exhibit different line shapes, and the emission features (σ and π) appear to be dispersive in nature. The emitted photon energies ($h\nu_{out}$) vary with the incident photon energy ($h\nu_{in}$) from 284.8 to 310.0 eV. The dashed lines in Figure 3 show these dispersive features. The $h\nu_{in}$-dependent emission features arise from transitions from states with a well-defined crystal momentum.[38,39] The π emission derives from transition from $p_z$ states that are oriented perpendicular to the graphitic plane. Emission from the σ bands is much more isotropic because these bands are derived in part from the $p_x$ and $p_y$ states that are parallel to the surface. The changes in the π feature are drastic for incident photon energies varied from 284.8 to 285.4 eV and then 288.4 eV, as shown in Figure 3. At an excitation energy of $h\nu_{in}$ = 284.8 eV, which is close to the energy of the π absorption features at ∼284 eV, the σ component of emission of HOPG has a distinct feature at ∼273 eV, whereas the π component of GNFs:N and GNFs is observed at ∼283.5 eV, and the σ component of the emission of GNFs:N and GNFs has a pronounced feature at ∼273.5 eV. As the excitation energy $h\nu_{in}$ is increased to 285.4 eV, the intensity of the π feature observed at ∼283.5 eV is decreased, whereas the most prominent feature of the σ component of GNFs:N and GNFs is observed at ∼274 eV. The σ feature is symmetric and the strongest in the $h\nu_{in}$= 288.4 eV spectrum. In this spectrum, the π feature is observed at ∼280 eV and becomes broader. The π and σ features are greatly broadened for $h\nu_{in}$= 310.0 eV because the states near the two critical points of the crystal can be reached by the excitation. The RIXS spectra in Figure 3 clearly provide evidence that the spectral intensities of GNFs:N are clearly larger than those of GNFs and HOPG for all excitation energies considered, confirming that the incorporation of N in the GNFs matrix increases the number of substitution-induced states; in particular, the DOSs in the π region and in the gap between π and π* are markedly increased for $h\nu_{in}$= 284.8 and 285.4 eV.

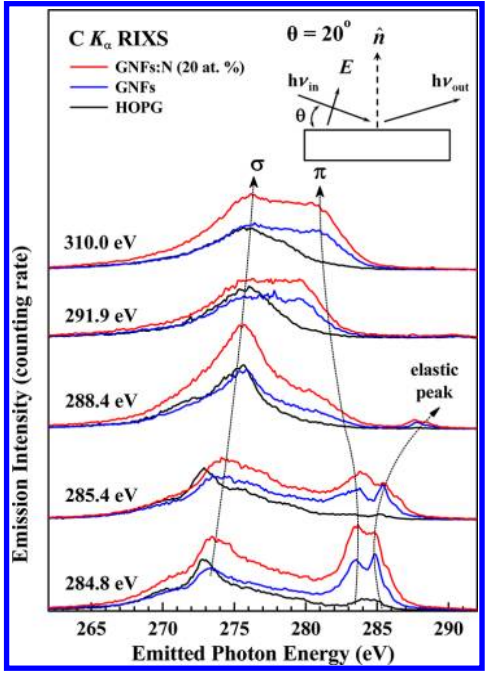

**Figure 3.** C $K_\alpha$ RIXS spectra of GNFs:N (20 at %), GNFs, and HOPG obtained with various incident photon energies. Upper inset displays experimental geometry.

Because the band gap of carbon-related materials depends strongly on the size, shape, and fraction of the $sp^2$ domains, PL emission can be tuned by controlling the nature of the $sp^2$ sites.[23] In this study, PL is also observed for GNFs, which contain graphitic C−N bonding or pyrrolic-N structure, as discussed in the Supporting Information. However, in the O K-edge XANES spectra of GNFs:N and GNFs (shown in the inset of Figure S4 in the Supporting Information), the observation of the π* at ∼533 eV and σ* at ∼541 eV, and the increased intensities of π* and σ* with an increased N content at 20 at % indicate the presence of the C=O bonding. These observations reveal that the character of PL of O- and N-treated GNFs depends strongly on the electronic structures and bonding properties. However, it has been suspected that the contaminated oxygen oxidized the GNFs, which disrupts the π network in the GNFs structure and generates a band gap when the graphene is composed of only a single/few sheets, as we observed in the O K-edge XANES and $K_\alpha$ XES of GO and reduced-GO that will be publish elsewhere. If the edges break the sublattice symmetry, then lateral confinement of the wave function may produce a band gap at its charge neutrality level. Therefore, the presence of oxygen makes the graphene layers luminescent by breaking some of the carbon−carbon bonds or causing functionalization (with nitrogen) at the edges of the sheets and forms N and O bonding that alters the electronic structure of the $sp^2$ clusters and their symmetry; as a result, PL extends across large areas of the layers[3,11,12] and hence contributes to the enhanced luminescence. However, as revealed by the RIXS spectra, the DOSs of GNFs:N and GNFs in the region of π and in the gap between π and π* are substantially enhanced, especially upon excitation by light with $h\nu_{in}$ = 284.8 and 285.4 eV. The presence of $sp^2$ domains in the carbon-(oxygen, nitrogen) $sp^3$ matrix can result in localization of electron−hole (e−h) pairs, facilitating radiative recombination in small/nanoscale clusters.[5,13] Meanwhile, the possible presence of epoxide and carboxylic groups may also induce nonradiative recombination of localized e−h pairs or give rise to a broad energy gap,[15] which leads to a relatively low and broad (the intensity of long-wavelength 650−800 nm) PL feature for GNFs:N and GNFs, as presented in Figure 1c. Furthermore, a study of PL of Mn-ion-bonded r-GO has suggested that the resonance energy can be transferred from the Mn ions to the $sp^2$ clusters, yielding intense long-wavelength emission.[15] To confirm that N and O atoms in the GNFs matrix can indeed enhance PL, XEOL of GNFs:N, and GNFs at the N, C and O K-edge, we performed excited photon energies and presented them in Figures 4a−h. XEOL is the emission of optical photons following the absorption of X-ray of a selected energy, often across an absorption edge. XEOL tracks the optical emission following X-ray excitation. It differs from PL in that it excites the core electrons of a specific atom in the chemical environment preferentially, that it has access to highly excited states, and that the thermalization of the photoelectrons/Auger electrons and holes in the system of interest primarily contributes to the luminescence. It has been demonstrated as a powerful tool for tracking the efficiency of a luminescence channel across absorption edges of atomic-site specificity, especially in the soft X-ray region.[43−45] One sees in Figure 4a−h a 3D plot of the excitation energy versus the optical emission with the intensity colored coded (z axis). The incident photon energies at the N and O K-edge yield predominately XEOL centered at ∼520 nm (Figures 4a,c,d,f,h), particularly at the O K-edge XEOL. These are compared with that at the C K-edge, where the XEOL intensity is relatively weak when the incident excited energies were ∼280−310 eV (Figures 4b,e,g), respectively. The emission wavelength (∼520 nm) of XEOL at the N and O K-edge, as shown in Figure 4, is close to that of the main PL feature (500−510 nm) of GNFs:N presented in Figure 1c. This finding is consistent with the previous reports[5,25] that the enhancement of the PL emission can be related to N and O or their ratio and further demonstrates that N and O atoms act as active luminescence centers and the resonance energy can be transferred from N and O atoms to the sp$^2$ clusters following the absorption of N and O K-edge X-ray excited energy, causing radiative relaxation in the π* and π gap, yielding intense short-wavelength emission, revealed by the main PL feature in Figure 1c. Figure 5 elucidates a plausible mechanism of PL in GNFs:N, which involves resonant transfer of energy from the N and O atoms to the $sp^2$ clusters. When GNFs:N were excited with suitable photon energy (say $E = h\nu$), then the core electrons from both N and O 1s states were excited to unoccupied 2p states that belonged to $\pi^*$ states ($sp^2$ antibonding states). After that, the electrons were intended to fall in $\pi$ states ($sp^2$ bonding states) or substitution-induced localized states in the region of $\pi$ or in the $\pi^*-\pi$ gap; then, the energy was released as the form of $E_{PL}$. As a result, the rate of radiative recombination in the gap between $\pi$ and $\pi^*$ was increased significantly because the DOSs in the $\pi$ region and the gap between $\pi$ and $\pi^*$ were increased, and the intensity of the PL feature was therefore enhanced. The transfer of energy from the N and O atoms to the $sp^2$ clusters in the GNFs matrix is highly feasible because the N and O K-edge XANES spectra reveal different nitrogen and oxygen chemical environments in the region of the $\pi^*$ feature, which are mainly associated with pyridine-like N–C, graphite-N-like, and the C=O bonding, as presented in the Supporting Information (Figure S4). Notably, GNFs:N contain more $sp^2$ clusters than GNFs, as observed in the XANES, XES, and RIXS spectra previously stated. The higher number of $sp^2$ clusters is responsible for the higher PL intensity for GNFs:N. Overall, as displayed in Figure 5, the general PL behavior of GNFs:N involves two mechanisms. The first involves the epoxide or carboxylic groups in GNFs, which induce nonradiative recombination of localized e–h pairs that leads to a low and broad PL, as revealed by the intensities of the spectra of GNF:N and GNFs. The second primarily involves radiative recombination, and resonant transfer of energy from the N and O bonding alters the electronic structure of the $sp^2$ clusters and their symmetry, which causes a blue shift and contributes to the enhanced the main PL emission feature of GNFs.

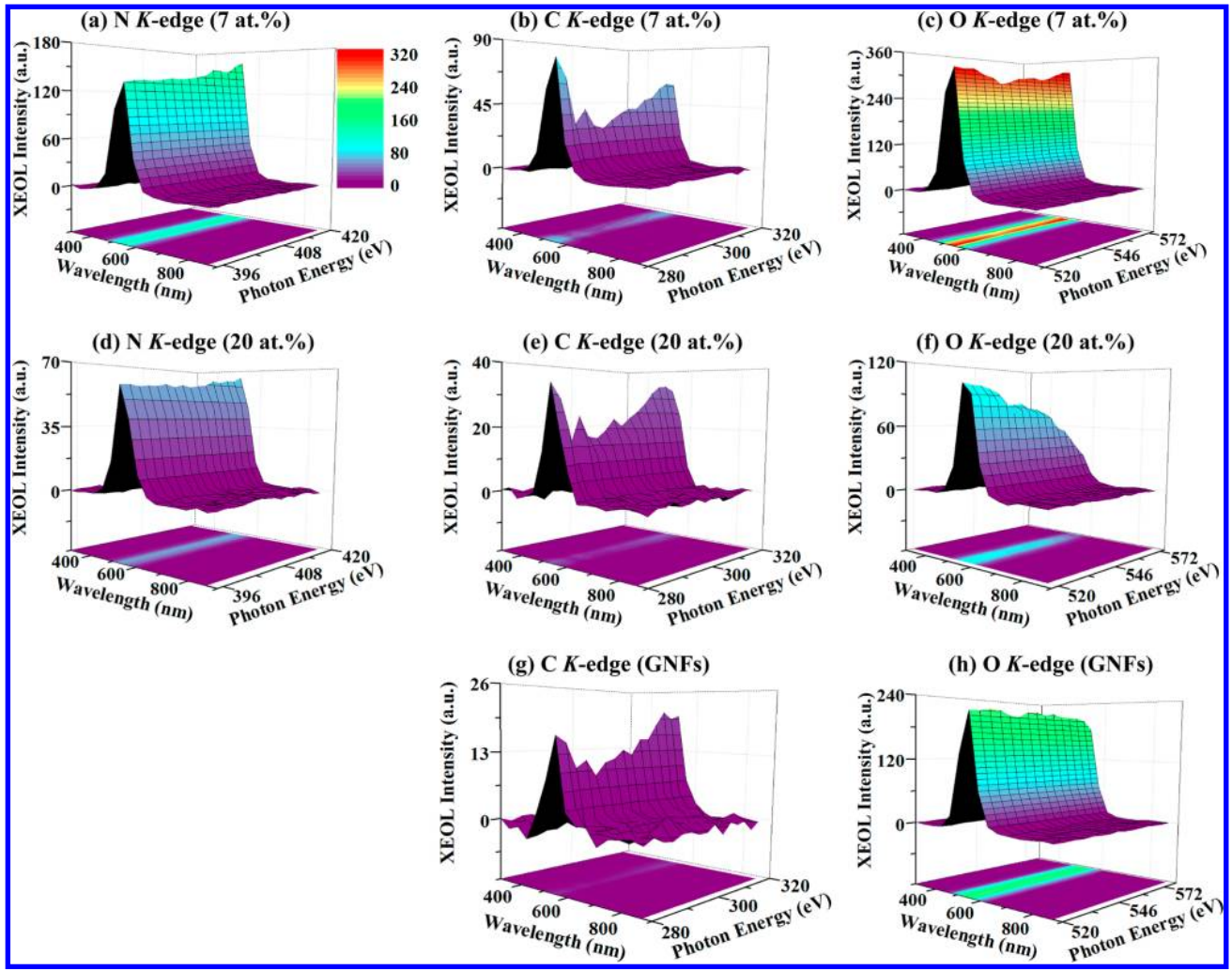

**Figure 4.** XEOL of GNFs:N and GNFs at the N K-edge (a,d); C K-edge (b,e,g); and O K-edge (c,f,h), respectively.

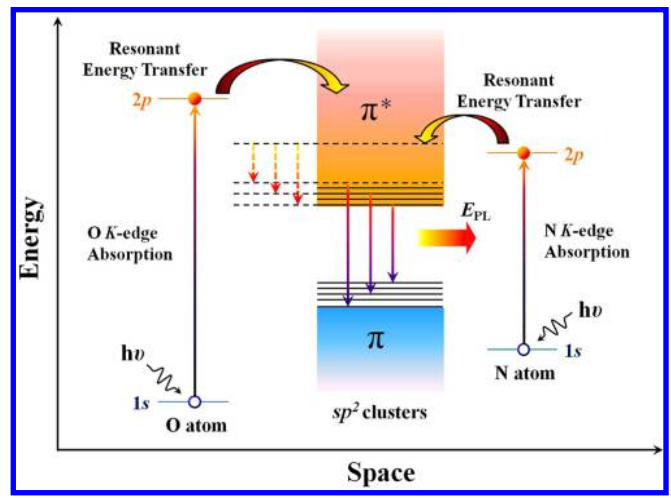

**Figure 5.** Schematic diagram of the possible mechanism of a large enhancement of PL by resonant energy transfer from the N and O atoms to the sp$^2$ clusters in the GNFs matrix. $E_{PL}$ stands for the enhanced PL emission feature. Solid and dotted lines represent radiative ($E_{PL}$) and nonradiative (low and broad PL) relaxation processes, respectively.

## 4. CONCLUSIONS

In conclusion, the PL behaviors of GNFs:N and GNFs are observed to depend strongly on the N- and O-bonded $sp^2$ clusters, epoxide, or carboxylic groups. The strong effect on the PL emission in GNFs:N and GNFs is related to the induced changes of the electronic structures and bonding properties, revealed by XANES, XES, and RIXS. XEOL spectra show strong emission at the N and O K-edge excited photon energy, which directly verifies that the N and O atoms rather than the C atoms play a critical role in main PL emission feature. The large increase in the intensity of the PL emission is caused by the resonant transfer of energy from the N and O atoms to the $sp^2$ clusters in the GNFs matrix. The XANES, XES, and RIXS results consistently show the same PL behavior. This study finds that the characteristic of the PL emission in GNFs can be tailored through functionalization with nitrogen/oxygen additions, which renders them attractive for novel optoelectronic devices.

## ■ ASSOCIATED CONTENT

### *S* Supporting Information

(C 1s, O 1s, N 1s X-ray photoemission spectroscopy and N K-edge, O K-edge X-ray absorption near-edge structure spectroscopy). This material is available free of charge via the Internet at http://pubs.acs.org.

## ■ AUTHOR INFORMATION

**Corresponding Author**
*E-mail: Sekhar.Ray@wits.ac.za (S.C.R.), jwchiou@nuk.edu.tw (J.W.C.), wfpong@mail.tku.edu.tw (W.F.P.). Tel:+27-011-7176806. Fax: +27-011-717-6879.

**Author Contributions**
[Φ]These authors contributed equally to this work

**Notes**
The authors declare no competing financial interest.

## ■ ACKNOWLEDGMENTS

J.W.C. and W.F.P. acknowledge the National Science Council of Taiwan for financial support under contract nos. NSC 99-2112-M390-004-MY3 and NSC99-2119-M032-004-MY3. S.C.R. also acknowledges National Research Foundation, South Africa for financial support.